\documentclass[preprint,11pt,aps,draft,nofootinbib,showpacs,showkeys]{revtex4}
\usepackage{amssymb}

\begin{document}

\title{Predictive statistical mechanics and macroscopic time evolution: hydrodynamics and entropy production}

\author{Domagoj Kui\'{c}}
\email{dkuic@pmfst.hr} 

\affiliation{University of Split, Faculty of Science, N. Tesle 12, 21000 Split, Croatia}

\date{May 27th, 2016}

\begin{abstract}
In the previous papers (Kui\'{c} et al. in Found Phys 42:319–339, 2012; Kui\'{c} in arXiv:1506.02622, 2015), it was demonstrated that applying the principle of maximum information entropy by maximizing the conditional information entropy, subject to the constraint given by the Liouville equation averaged over the phase space, leads to a definition of the rate of entropy change for closed Hamiltonian systems without any additional assumptions. Here, we generalize this basic model and, with the introduction of the additional constraints which are equivalent to the hydrodynamic continuity equations, show that the results obtained are consistent with the known results from the nonequilibrium statistical mechanics and thermodynamics of irreversible processes. In this way, as a part of the approach developed in this paper, the rate of entropy change and entropy production density for the classical Hamiltonian fluid are obtained. The results obtained suggest the general applicability of the foundational principles of predictive statistical mechanics and their importance for the theory of irreversibility.
\end{abstract}

\keywords{maximum entropy principle, statistical mechanics, nonequilibrium theory, Hamiltonian dynamics, hydrodynamics, thermodynamics of irreversible processes, entropy production}

\maketitle

\section{Introduction}\label{eqI}

Here we continue the study started in \cite{kuic,kuic1} of the application of predictive statistical mechanics to a problem of predicting the macroscopic time evolution of systems with Hamiltonian dynamics, in the case when the information about the microscopic dynamics is not complete. For this purpose, in \cite{kuic} we developed the basic theoretical model for closed systems with time independent Hamiltonian function, and it was generalized in \cite{kuic1} to include also closed systems with the Hamiltonian function that depends on time. Furthermore, in \cite{kuic,kuic1} we also gave a brief introduction about the Shannon's \cite{shannon} concept of information entropy as the measure of uncertainty represented by the probability distribution, and also on the principles of maximum information entropy and macroscopic reproducibility, which are the foundational principles of predictive statistical mechanics formulated by E. T. Jaynes \cite{jaynes1,jaynes2,jaynes3,jaynes5,jaynes4,jaynes6,jaynes7,jaynes8}.  Here we only mention that the principle of maximum information entropy represents the general criterion for the construction of probability distribution when the available information is not sufficient the unique determination of the distribution \cite{jaynes1,jaynes2}. Maximization of the information entropy subject to given constraints is an algorithm of the construction of the probability distribution (MaxEnt), such that only the information represented by these constraints is included in the probability distribution. If, by controling certain macroscopic quantities,  the same macrosopic behavior is reproduced in the experiment, then, according to the principle of macroscopic reproducibility, the information about the values of those quantities is relevant for the prediction of that macroscopic phenomena. Sharp, definite predictions of macroscopic behavior are possible only because certain behavior is characteristic for each of the overwhelming majority of microstates compatible with data, and therefore, this is just the behavior that is reproduced experimentally under those constraints. That was confirmed also by the conclusions reached in the framework of MaxEnt formalism by Grandy \cite{grandy1,grandy2,grandy3,grandy4,grandy}.

In the interpretation given by Jaynes, irreversibility of physical processes reflects only our inability to follow the exact state of the system during the process, and it can be considered a consequence of the associated loss of information as to the state of the system \cite{jaynes2}. In our previous papers \cite{kuic,kuic1} we have demonstrated that such interpretation has a clear mathematical formulation in the concept of maximization of the conditional information entropy and its relation with the information entropy. In the basic model presented in \cite{kuic,kuic1}, this resulted in the definition of the entropy change and the rate of entropy change for a closed Hamiltonian system without additional assumptions. This paper is devoted to the generalization of the approach developed in \cite{kuic,kuic1}. Through the comparison with the reduced description of nonequilibrium systems from reference \cite{zubarev}, we have concluded that for a such generalization, the data about the quantities relevant for the prediction on the specified time scales should be included in the basic theoretical model. According to \cite{zubarev}, if we are interested in behavior of the system for time intervals that are not too small in the specified sense, the details of the initial state become unimportant and the number of parameters necessary for the description of the state of the system is reduced. Other methods that use reduced descriptions of nonequilibrium states along with the quantum or classical Liouville equations are also known from the literature \cite{zwanzig1,zwanzig2,robertson1,robertson2,zubarev2}. We have selected the hydrodynamic time scale for the first step in the generalization of our approach, where for the description of a nonequilibrium system less detailed information about the microscopic dynamics is required in comparison to other time scales. 

Since the hydrodynamic continuity equations represent the basic element of a reduced description on the hydrodynamic time scale, in Section \ref{secHJK} they are taken as the relevant information which, as additional constraints on the maximization of the conditional information entropy, should be included in our initial model. The equivalent form of these equations, suitable for the use in the variational calculation is also derived. In Section \ref{MRATHTS} conditions are verified under which this equivalence holds and the form of the hydrodynamic continuity equations does not depend on the missing information about the microscopic dynamics. Predictions that follow from the maximization of the conditional information entropy, subject to the constraint given by the Liouville equation averaged over the available phase space, and additional constraints that are equivalent to the hydrodynamic continuity equations, are derived in Section \ref{secMIHIVE}. We show there that this generalized approach results in the microstate probability distribution which is identical in form to the relevant distribution for the classical fluid in local equilibrium known from the literature \cite{zubarev}. Furthermore, the expression for the rate of entropy change is obtained in accordance with the corresponding expression known from the thermodynamics of irreversible processes \cite{evans,deGroot}. This allows us to define the density of entropy production consistently with its basic postulates. In Section \ref{cswef} corresponding results are also obtained for the classical fluid with external forcing. By further generalization of this approach in the further paper, open systems in contact with particle reservoirs will be discussed, and the transport coefficients for the classical fluid will be derived accordingly.

\section{Hydrodynamic continuity equations} \label{secHJK}

Following ref. \cite{zubarev}, for the time intervals longer than the time  $\tau _r$ for the establishment of local equilibrium, local macroscopic quantities, such as  the local particle-number density, the local momentum density and the local energy density, are sufficient to describe the nonequilibrium system. For the classical fluid of $N$ identical particles, taken here as the basis for the analysis, the dynamical variables that correspond to these quantities are the particle-number density
\begin{eqnarray}
n ({\bf r}) \equiv n ({\bf r}; {\bf r}_1, \dots, {\bf r}_N) = \sum_{i=1}^N \delta ({\bf r} - {\bf r}_i)\ , \label{eq78}
\end{eqnarray}
momentum density
\begin{eqnarray}
{\bf P}({\bf r}) \equiv {\bf P}({\bf r}; {\bf r}_1, {\bf p}_1,\dots, {\bf r}_N, {\bf p}_N) = \sum_{i=1}^N {\bf p}_i \delta ({\bf r} - {\bf r}_i)\ , \label{eq79}
\end{eqnarray}
and energy density
\begin{eqnarray}
h({\bf r}) & \equiv & h({\bf r}; {\bf r}_1, {\bf p}_1,\dots, {\bf r}_N, {\bf p}_N) \nonumber\\
& = & \sum_{i=1}^N  \left [\frac{{\bf p}_i^2}{2m} + \frac{1}{2}\sum_{j=1,\ j\neq i}^N  \Phi (\vert {\bf r}_i - {\bf r}_j \vert)\right ] \delta ({\bf r} - {\bf r}_i)\ . \label{eq80}
\end{eqnarray}
The classical fluid of $N$ identical particles is described by the translationaly and rotationaly invariant Hamiltonian function:
\begin{equation}
H(x, p) = \sum_{i=1}^N  \left [\frac{{\bf p}_i^2}{2m} + \frac{1}{2}\sum_{j=1,\ j\neq i}^N  \Phi (\vert {\bf r}_i - {\bf r}_j \vert)\right ]\ , \label{eq81}
\end{equation}
where $\Phi (\vert {\bf r}_i - {\bf r}_j \vert)$ is the potential energy of interaction of  the particle pair with indices  $i, j$. The set of $6N$ dynamical variables is denoted by $(x,p)$ and it consists of the Cartesian components of $N$ position vectors $({\bf r}_1, \dots, {\bf r}_N)$ and corresponding $N$ momentum vectors $({\bf p}_1, \dots, {\bf p}_N)$. Time dependence of the variables $(x,p)$ is determined by Hamilton's equations
\begin{equation}
\dot x_i = \frac{\partial H}{\partial p_i} , \qquad \qquad \dot p_i =  - \frac{\partial H}{\partial x_i} , \qquad \qquad1 \leq i \leq 3N\ , \label{eq82}
\end{equation} 

It is easy to show that the integrals of the dynamical variables (\ref{eq78}), (\ref{eq79}) and (\ref{eq80}) taken over the entire volume of the system give the total particle number $N$, the total momentum ${\bf P}_{tot}$ and the Hamiltonian function $H(x, p)$ given by (\ref{eq81}), respectively:
\begin{eqnarray}
N & = & \int n ({\bf r}; {\bf r}_1, \dots, {\bf r}_N)\, d^3{\bf r} \ , \label{eq88} \\ 
{\bf P}_{tot} & = & \sum_{i=1}^N {\bf p}_i = \int {\bf P}({\bf r}; {\bf r}_1, {\bf p}_1,\dots, {\bf r}_N, {\bf p}_N)\, d^3{\bf r} \ , \label{eq89} \\
H(x, p) & = & \int h({\bf r}; {\bf r}_1, {\bf p}_1,\dots, {\bf r}_N, {\bf p}_N)\, d^3{\bf r}\ . \label{eq87}
\end{eqnarray}
The total particle number $N$ is fixed and the Hamiltonian function $H(x, p)$ given by (\ref{eq81}) is time independent. The same is true for the total momentum ${\bf P}_{tot}$, given by (\ref{eq89}), because the Hamiltonian function $H(x, p)$, given by (\ref{eq81}), is translationaly invariant. This can be expressed in terms of Poisson brackets:
\begin{eqnarray}  
\frac{dH}{dt} & = & \{H, H\} = 0\ , \label{eq95} \\
\frac{d{\bf P}_{tot}}{dt} & = & \{{\bf P}_{tot},H\} = 0\ . \label{eq94} 
\end{eqnarray}

Local values of the macroscopic quantities $\langle n ({\bf r})\rangle _t$, $\langle {\bf P}({\bf r}) \rangle _t$, $\langle h({\bf r}) \rangle _t$ that describe the classical fluid of identical particles are obtained by averaging the dynamical variables $n ({\bf r})$, ${\bf P}({\bf r})$ i $h({\bf r})$ over the microstate probability density function $f(x,p,t)$ at time $t$:
\begin{eqnarray}
\langle n ({\bf r})\rangle _t & = & \int_M   f (x, p, t) n ({\bf r}; {\bf r}_1, \dots, {\bf r}_N)\, d\Gamma \ , \label {90} \\
\langle {\bf P}({\bf r}) \rangle _t & = & \int_M f (x, p, t)\, {\bf P}({\bf r}; {\bf r}_1, {\bf p}_1,\dots, {\bf r}_N, {\bf p}_N)\, d\Gamma \ , \label {91}\\
\langle h({\bf r}) \rangle _t & = & \int_M f (x, p, t)\, h({\bf r}; {\bf r}_1, {\bf p}_1,\dots, {\bf r}_N, {\bf p}_N)\, d\Gamma \ , \label {92}
\end{eqnarray}
where $d\Gamma = dx_1\dots dx_{3N} dp_1\dots dp_{3N}$ is the volume element of the $6N$-dimensional phase space $\Gamma $. Averages (\ref {90}), (\ref {91}) and (\ref {92}) are given by the integrals over the set $M\subset \Gamma $ which corresponds to all possible microstates. The set $M$ is that part of the phase space $\Gamma $ which is accessible to the system. By definition the set $M$ is taken here to be invariant to Hamiltonian motion of points $(x,p)$ in $\Gamma $, as determined by the equations of motion (\ref{eq82}). The boundaries of the set $M\subset \Gamma $ are determined by the conservation laws given by (\ref{eq95}) and (\ref{eq94}), along with other possible conservation laws for the Hamiltonian function (\ref{eq81}), and by our prior information on the possible values of these conserved quantities. Thus, together this determines the sample space, given by the set $M$, on which the probability density function $f(x,p,t)$ is defined. Properties of the set $M$ will be further elaborated in Section \ref{MRATHTS}. The microstate probability density function $f(x,p,t)$ for the system of $N$ identical particles is normalized in accordance with the definition of microstates in the phase space that follows in the classical limit of quantum statistical mechanics \cite{zubarev}.

The local dynamical variables $n ({\bf r})$, ${\bf P}({\bf r})$ and $h({\bf r})$ are the densities of the corresponding conserved quantities $N$, ${\bf P}_{tot}$ i $H(x,p)$. The equations of motion of these dynamical variables can therefore be written in the form of the local microscopic conservation laws \cite{zubarev},
\begin{eqnarray}  
\frac{d n ({\bf r})}{dt} & = & \{n ({\bf r}), H\}  =  - \nabla \cdot {\bf J} ({\bf r}) \ , \nonumber \\
\frac{d P_\alpha  ({\bf r})}{dt} & = & \{P_\alpha ({\bf r}), H\}  =  - \nabla \cdot {\bf J}_{P_\alpha}({\bf r})\ , \nonumber \\
\frac{d h ({\bf r})}{dt} & = & \{h ({\bf r}), H\}  =  - \nabla \cdot {\bf J}_h ({\bf r}) \ . \label{eq99}
\end{eqnarray}
where $\{ , \}$ denotes the Poisson brackets. The Cartesian components of the momentum density vector ${\bf P}({\bf r})$ are denoted by $P_\alpha ({\bf r})$, $\alpha = 1, 2, 3$. The dynamical variables ${\bf J} ({\bf r})$, ${\bf J}_{P_\alpha}({\bf r})$ and ${\bf J}_h ({\bf r})$ are the flux densities of conserved quantities whose densities are $n ({\bf r})$, $P_\alpha ({\bf r})$ i $h({\bf r})$. Equations (\ref{eq99}) are standard expressions; explicit derivations of the flux densities are found in the literature \cite{grandy,zubarev,evans,balian}.

The average value of the time derivative of  any dynamical variable $A$ is equal to the time derivative of the average value $\langle A \rangle _t = \int_M Af\, d\Gamma $ of the same variable: 
\begin{eqnarray}
\left \langle {d A \over d t} \right \rangle_t & = & 
\int_M \left ({\partial  A \over \partial  t} f +  \{A, H\} f\right )\, d\Gamma \nonumber\\ 
& = & \int_M \left ({\partial  A \over \partial  t} f -  A \{f, H\} \right )\, d\Gamma \nonumber\\
& = & \int_M \left ({\partial  A \over \partial  t} f +  A{\partial  f \over \partial  t} \right )\, d\Gamma \nonumber\\
& = & {d \langle A \rangle _t \over d t} \ . \label{eq100}
\end{eqnarray}
The derivation of (\ref{eq100}) uses: the equation of motion for the dynamical variable $A$,
\begin{equation}
{d A \over d t} = {\partial  A \over \partial  t} + \{A, H\}\ ,  \label{eq101}
\end{equation}
the Liouville equation for the microstate probability density $f(x,p,t)$,
\begin{equation}
{d f \over d t} = {\partial  f \over \partial  t} + \{f, H\} = 0 \ ,  \label{eq102}
\end{equation}
and the $n$-dimensional generalization of the divergence theorem \cite{ostrogradski}, the application of which along with the vanishing of the contribution of the boundary of the invariant set $M$ (the explanation is analogous to that given for (\ref{eq123d})) gives the second line of (\ref{eq100}).

By averaging (\ref{eq99}) over the microstate probability density $f(x,p,t)$ and using (\ref{eq100}) one obtains the following expressions:
\begin{eqnarray}  
\frac{\partial  \langle n ({\bf r})\rangle_t }{\partial t} & = & \langle \{n ({\bf r}), H\} \rangle_t = - \nabla \cdot \langle {\bf J} ({\bf r}) \rangle_t \ ,  \nonumber \\
\frac{\partial \langle P_\alpha  ({\bf r}) \rangle_t}{\partial t}   & = & \langle \{P_\alpha ({\bf r}), H\} \rangle_t =  - \nabla \cdot \langle {\bf J}_{P_\alpha}({\bf r}) \rangle_t \ ,  \nonumber \\
\frac{\partial \langle h ({\bf r})\rangle_t }{\partial t}    & = &  \langle \{h ({\bf r}), H\} \rangle_t =  - \nabla \cdot \langle {\bf J}_h ({\bf r}) \rangle_t \ . \label{eq105}
\end{eqnarray}
Time derivatives in (\ref{eq105}) are denoted as partial derivatives because the average values of the densities depend on the position vector ${\bf r}$ also. Equations (\ref{eq105}) are the local macroscopic conservation laws, which serve the basis  for derivation of the  hydrodynamic equations \cite{grandy,zubarev,evans,balian}.

An important step in the  derivation of equality (\ref{eq100}) for an arbitrary dynamical variable $A$, and then also in the derivation of (\ref{eq105}), was the use of the Liouville equation (\ref{eq102}). Furthermore, it can be shown directly that the equations  
\begin{eqnarray}
\int_M n ({\bf r}) \left( {\partial  f \over \partial  t} + \{f, H\} \right ) d\Gamma & = & 0 \ , \nonumber \\
\int_M {\bf P}({\bf r}) \left( {\partial  f \over \partial  t} + \{f, H\} \right ) d\Gamma & = & 0 \ , \nonumber \\
\int_M h({\bf r}) \left( {\partial  f \over \partial  t} + \{f, H\} \right ) d\Gamma & = & 0 \ , \label{eq108}
\end{eqnarray}
are equivalent to the local macroscopic conservation laws (\ref{eq105}). By using the divergence theorem in expressions (\ref{eq108}) along with the vanishing of the contribution of the boundary of the set $M$, in the way described in the derivation of (\ref{eq100}), and then using the right hand side of (\ref{eq99}), we obtain the equivalent expressions
\begin{eqnarray}
& &\int_M \left( {\partial  f \over \partial  t} n ({\bf r}) + f\nabla \cdot {\bf J} ({\bf r}) \right ) d\Gamma = 0 \ , \nonumber \\
& &\int_M \left( {\partial  f \over \partial  t} {\bf P}({\bf r}) + f\nabla \cdot {\bf J}_{P_\alpha}({\bf r}) \right ) d\Gamma = 0 \ , \nonumber \\
& &\int_M \left( {\partial  f \over \partial  t} h({\bf r}) + f\nabla \cdot {\bf J}_h ({\bf r}) \right ) d\Gamma = 0 \ . \label{eq111}
\end{eqnarray}
These expressions are the local macroscopic conservation laws (\ref{eq105}). 

\section{Macroscopic reproducibility and the hydrodynamic time scale} \label{MRATHTS}

Let us assume now, that along with the Hamiltonian function $H(x,p)$ given by (\ref{eq81}), the total Hamiltonian function $H_{tot}(x,p,t)$ includes also an additional term $H_{ni} (x,p,t)$, about which we do not have any prior information,
\begin{equation}
H_{tot}(x,p,t) = H(x,p) + H_{ni}(x,p,t)\ . \label{eq112}
\end{equation}
Let us assume now that some microstate probability density function $\tilde f(x,p,t)$ really satisfies ``the total'' Liouville equation
\begin{eqnarray}
{\partial  \tilde f \over \partial  t} + \{\tilde f, H_{tot}\} = {\partial  \tilde f \over \partial  t} + \{\tilde f, H\} +  \{\tilde f, H_{ni}\} = 0 \ .  \label{eq113}
\end{eqnarray}
As an addition, let us assume that the invariant set $M$ of all possible microstates in the phase space, is invariant also on the time evolution described by the total Hamiltonian function $H_{tot}(x,p,t)$. This situation can be imagined in a case where the set of dynamical variables exists that are constants of motion for both Hamiltonian functions, $H(x,p)$ and $H_{tot}(x,p,t)$. If such an assumption is unrealistic, we can assume instead the much simpler possibility that the invariant set $M$ is the entire phase space $M = \Gamma $.

Under these three assumptions the following statements are true. If the equations 
\begin{eqnarray}  
\int_M \tilde f\{n ({\bf r}), H_{ni}\}\, d\Gamma  & = & 0 \ , \nonumber \\
\int_M \tilde f\{P_\alpha ({\bf r}), H_{ni}\}\, d\Gamma  & = & 0 \ , \nonumber \\
\int_M \tilde f\{h ({\bf r}), H_{ni}\}\, d\Gamma  & = & 0 \ , \label{eq115}
\end{eqnarray}
are satisfied then the local macroscopic conservation laws are valid in the form which is identical to (\ref{eq105}). 

If the equations
\begin{eqnarray}  
\{n ({\bf r}), H_{ni}\}  & = & 0 \ , \nonumber \\
\{P_\alpha ({\bf r}), H_{ni}\}  & = & 0 \ , \nonumber \\
\{h ({\bf r}), H_{ni}\}  & = & 0 \ , \label{eq114}
\end{eqnarray}
are satisfied then the local microscopic conservation laws are valid in the form which is identical to (\ref{eq99}). The condition (\ref{eq114}) is more restrictive for $H_{ni} (x,p,t)$ than (\ref{eq115}); if the condition (\ref{eq114}) is satisfied then also the condition (\ref{eq115}) is satisfied. 

Previous statements can essentially be summarized in the following way. If ``the total'' Liouville equation (\ref{eq113}) is valid, and if the condition (\ref{eq114}) or condition (\ref{eq115}) is satisfied, then the local macroscopic conservation laws are still valid in the same form (\ref{eq105}), which is equivalent to equations (\ref{eq108}). 

It is important also that (\ref{eq115}) and (\ref{eq114}) can not be used in predictions with the help of the maximum entropy principle, because we do not have prior information about the term $H_{ni}(x,p,t)$ of the total Hamiltonian function $H_{tot}(x,p,t)$. It is also important to note the following: if some function $\tilde f(x,p,t)$ satisfies the total Liouville equation (\ref{eq113}) and equations (\ref{eq115}), then this function also satisfies the equations (\ref{eq108}). The logical converse is not valid.

Equations (\ref{eq115}) and (\ref{eq114}) can be interpreted in the following way. Equations (\ref{eq114}) are statements that the missing information  about the microscopic dynamics is not relevant for the description of time evolution of the local dynamical variables $n ({\bf r})$, $P_\alpha ({\bf r})$ and $h ({\bf r})$. Equations (\ref{eq115}) are statements that the missing information  about microscopic dynamics is not relevant for the description of  time evolution of the local macroscopic quantities $\langle n ({\bf r})\rangle _t$, $\langle P_\alpha ({\bf r})\rangle _t$ and $\langle h ({\bf r})\rangle _t$. Both statements are in accordance  with the assumption that the reduced description of nonequilibrium macroscopic systems is possible on the specified time scales, as was discussed in detail in the previous section. That assumption can be accepted as the consequence of the foundational principle of macroscopic reproducibility.

Thus, condition (\ref{eq115}) or (\ref{eq114}) is verified under which equations (\ref{eq108})  are equivalent in form to the macroscopic continuity equations. These are the conditions under which the missing information about the microscopic dynamics (i.e. missing information about the term $H_{ni}$ of the total Hamiltonian function $H_{tot}$ in (\ref{eq112})) is not important for the form of the macroscopic continuity equations (\ref{eq105}). It is explained why these conditions are so important for the reduced description of the system. The macroscopic continuity equations allow a further derivation of the hydrodynamic equations; they are the basic elements of the reduced description of the macroscopic time evolution on the hydrodynamic time scale. As a consequence, under condition (\ref{eq115}) or (\ref{eq114}), missing information about the microscopic dynamics is not relevant for a reduced description of a nonequilibrium system at a hydrodynamic time scale.

\section{MaxEnt and hydrodynamic irreversible time evolution}\label{secMIHIVE}

In relation to the basic model developed in the previous papers \cite{kuic,kuic1}, the macroscopic conservation laws (\ref{eq105}) represent the relevant additional information that is foundational for the description of nonequilibrium system on the hydrodynamic time scale. In the basic model, the only constraints on the maximization of the conditional information entropy, 
\begin{eqnarray}
S_{I}^{DF} (t_a, t_0) & = & - \int _{S_0(M)} \int_\Gamma DF\log D \ d\Gamma dS _0 \nonumber \\
& = & - \int _{t_0}^{t_a} \int _{S_0(M)} \int_M \frac{\partial D}{\partial t}F\log D \ d\Gamma dS _0 dt + S_{I}^{DF}(t_0, t_0) , \label{eq115a}
\end{eqnarray}
were the normalization of the conditional probability density $D\equiv D(x,p,t | (x_0,p_0)_\omega ,t_0 )$,
\begin{eqnarray}
\varphi_1 ((x_0,p_0)_\omega ,t_0; t, D) = F \int_M D \ d\Gamma   -  F = 0 , \label{eq116}
\end{eqnarray} 
and the Liouville equation for $D(x,p,t | (x_0,p_0)_\omega ,t_0 )$ averaged over the available phase space (i.e. over the set $M \subset \Gamma $ of all possible microstates which is invariant to the Hamiltonian motion),
\begin{equation}
\varphi _2 ((x_0,p_0)_\omega ,t_0; t, D) =  \int_M \left [{\partial D \over \partial t} + \sum _{i=1} ^{3N} \left ({\partial D \over \partial x_i}{\partial H \over \partial p_i} - {\partial D \over \partial p_i}{\partial H \over \partial x_i}\right )\right ]F \ d\Gamma = 0 . \label{eq116a}
\end{equation}
The conditional probability density $D(x,p,t | (x_0,p_0)_\omega ,t_0 )$ corresponds to the conditional probability that at time $t$ the point corresponding to the state of the system is in the element $d\Gamma $ around $(x,p)$, if at time $t_0$ it is anywhere along the paths passing through the infinitesimal surface element $dS_0$ located at $(x_0,p_0)$ on the surface $S_0(M)$. A phase space path is uniquely determined by the solution of Hamilton's equations (\ref{eq82}). By definition, the surface $S_0(M)$ is perpendicular to all paths in the set $\Omega (M)$ of all phase space paths in $M$. The correspondence between the points  $(x_0,p_0)_\omega  \in S_0(M) $ and paths $\omega \in \Omega (M)$ is one-to-one and the measure defined on the surface $S_0(M) $ is utilized as the measure on the set $\Omega (M)$ of all phase space paths in $M $. The conditional probability density $D(x,p,t | (x_0,p_0)_\omega ,t_0 )$ is defined by
\begin{equation}
D(x,p , t \vert (x_0,p_0)_\omega , t_0) = {G(x,p,t;(x_0,p_0)_\omega ,t_0) \over F((x_0,p_0)_\omega ,t_0) } . \label{eq118} 
\end{equation}     
Here, $G(x,p,t;(x_0,p_0)_\omega ,t_0)$ is a joint probability density of two continuous multidimensional variables, $(x,p)$ in $\Gamma $ and $(x_0,p_0)_\omega$ in $S_0(M)$. Path probability density $F((x_0,p_0)_\omega ,t_0)$ is given by the integral
\begin{equation}
F((x_0,p_0)_\omega ,t_0) = \int_\Gamma  G(x,p,t; (x_0,p_0)_\omega ,t_0 )d\Gamma , \label{eq119}
\end{equation} 
Similarly, the microstate probability density is given by 
\begin{equation}
f(x,p, t) = \int _{S_0(M)}G(x,p,t; (x_0,p_0)_\omega ,t_0 )dS_0 .  \label{eq120}
\end{equation}
If the Hamilton's equations are time dependent then phase space paths are time dependent objects also. Therefore,  as explained in the previous paper \cite{kuic1}, the interpretation given to the function $D(x,p,t | (x_0,p_0)_\omega ,t_0 )$ is in the case of time dependent Hamilton's equations taken by the conditional probability density $B(x,p , t \vert x_0,p_0, t_0)$ defined by 
\begin{equation}
B(x,p , t \vert x_0,p_0, t_0) = {\mathcal{F}(x,p,t; x_0,p_0 ,t_0) \over f(x_0,p_0,t_0) } . \label{eq121}
\end{equation} 
Here, $\mathcal {F}(x,p,t;x_0,p_0 ,t_0)$ is the probability density function defined on the $4s$-dimensional space $\Gamma \times \Gamma $. The conditional information entropy  $S_{I}^{Bf}(t, t_0)$ is obtained by replacement of the symbols with corresponding meanings in (\ref{eq115a}) as explained in \cite{kuic1}: replace $G(x,p,t;(x_0,p_0)_\omega ,t_0)$ with $\mathcal{F}(x,p,t; x_0,p_0 ,t_0)$, $F((x_0,p_0)_\omega ,t_0 )$ with $f(x_0,p_0, t_0)$, $D(x,p , t \vert (x_0,p_0)_\omega , t_0)$ with $B(x,p , t \vert x_0,p_0 , t_0)$, $M $ and $S_0(M)$ with $\Gamma $, and $dS_0$ with $d\Gamma _0$.  With these replacements applied to (\ref{eq116}) and (\ref{eq116a}) we obtain the normalization condition and the Liouville equation for $B(x,p,t | (x_0,p_0)_\omega ,t_0 )$ averaged over the available phase space, respectively. Analogous replacements are then applied also in the rest of the paper.

The generalization of our approach \cite{kuic,kuic1} that will be exposed here includes both constraints (\ref{eq116}) and (\ref{eq116a}). The only difference with respect to the basic model are the additional constraints (\ref{eq108}) written here in the form
\begin{eqnarray}
&& \varphi _{n}({\bf r}, t, D) = \cr\nonumber\\
&& =  \int_M  \int_{S_0(M)} n({\bf r})\left [{\partial D \over \partial t} + \sum _{i=1} ^{3N} \left ({\partial D \over \partial x_i}{\partial H \over \partial p_i} - {\partial D \over \partial p_i} {\partial H \over \partial x_i}\right )\right ]F\, dS_0\, d\Gamma = 0 \ , \cr\nonumber\\
&& \varphi _{P_\alpha }({\bf r}, t, D)  =  \cr\nonumber\\
&& = \int_M  \int_{S_0(M)} P_\alpha ({\bf r})\left [{\partial D \over \partial t} + \sum _{i=1} ^{3N} \left ({\partial D \over \partial x_i}{\partial H \over \partial p_i} - {\partial D \over \partial p_i} {\partial H \over \partial x_i}\right )\right ]F\, dS_0\, d\Gamma = 0 \ , \cr\nonumber\\
&& \varphi _{h}({\bf r}, t, D)  =  \cr\nonumber\\
&& = \int_M  \int_{S_0(M)} h({\bf r})\left [{\partial D \over \partial t} + \sum _{i=1} ^{3N} \left ({\partial D \over \partial x_i}{\partial H \over \partial p_i} - {\partial D \over \partial p_i} {\partial H \over \partial x_i}\right )\right ]F\, dS_0\, d\Gamma = 0 \ , \label{eq116b}
\end{eqnarray}
where the index $\alpha = 1, 2, 3$ denotes the Cartesian components of the vector ${\bf P}({\bf r})$. In the variational problem  the additional constraints (\ref{eq116b}) are introduced with the help of the corresponding additional Lagrange multipliers $\lambda _n({\bf r}, t)$, $\lambda _{P_\alpha}({\bf r}, t)$, $\lambda _h({\bf r}, t)$ and the appropriate functionals
\begin{eqnarray}
C_n[D, \lambda _n] & = & \int _{t_0}^{t_a} \int \lambda _n({\bf r}, t)\, \varphi _n({\bf r}, t, D)\, d^3{\bf r}\, dt \ , \nonumber\\
C_{P_\alpha}[D, \lambda _{P_\alpha}] & = & \int _{t_0}^{t_a} \int \lambda _{P_\alpha}({\bf r}, t)\, \varphi _{P_\alpha}({\bf r}, t, D)\, d^3{\bf r}\, dt \ , \nonumber\\
C_h[D, \lambda _h] & = & \int _{t_0}^{t_a} \int \lambda _h({\bf r}, t)\, \varphi _h({\bf r}, t, D)\, d^3{\bf r}\, dt \ , \label{eq116c}
\end{eqnarray}
where $\varphi _n({\bf r}, t, D)$, $\varphi _{P_\alpha}({\bf r}, t, D)$ ($\alpha = 1, 2, 3$) and $\varphi _h({\bf r}, t, D)$ are the constraints given by (\ref{eq116b}). Similarly, constraints (\ref{eq116}) and (\ref{eq116a}) are introduced using the Lagrange multipliers $\lambda _1 ((x_0,p_0)_\omega ,t_0; t)$ and $\lambda _2((x_0,p_0)_\omega ,t_0; t)$ and the functionals:      
\begin{equation}
C_1[D, \lambda _1] = \int_{S_0(M)} \int _{t_0}^{t_a} \lambda _1 ((x_0,p_0)_\omega ,t_0; t) \varphi_1 ((x_0,p_0)_\omega ,t_0; t, D) \ dt dS_0  , \label{eq116d}
\end{equation} 
and
\begin{equation}
C_2[D, \lambda _2] = \int_{S_0(M)} \int _{t_0}^{t_a} \lambda _2((x_0,p_0)_\omega ,t_0; t) \varphi _2 ((x_0,p_0)_\omega ,t_0; t, D) \ dt dS_0. \label{eq116e}
\end{equation} 

It is suitable to form the following functional 
\begin{equation}
J[D] = S_{I}^{DF} (t_a, t_0) - S_{I}^{DF}(t_0, t_0) =  \int _{t_0}^{t_a} \int _{S_0(M)} \int_M  K(D, \partial _t D) d\Gamma dS _0 dt , \label{eq116f}
\end{equation}
with the function $K(D, \partial _t D)$ given by
\begin{equation}
K(D, \partial _t D) = - \frac{\partial D}{\partial t}F \log D . \label{eq116g}
\end{equation} 
The functional $J[D]$ in (\ref{eq116f}) is rendered here stationary with respect to variations subject to the constraints (\ref{eq116}), (\ref{eq116a}) and (\ref{eq116b}). As explained in the previous papers \cite{kuic,kuic1}, the Euler equation for the constraint (\ref{eq116a}) is equal to zero, and we apply the most general multiplier rule for this type of problems taken from ref. \cite{wan} by introducing an additional constant Lagrange multiplier $\lambda _0$ for the function $K$, 
\begin{equation} 
J[D, \lambda_0] = \int _{t_0}^{t_a} \int _{S_0(M)} \int_M  \lambda _0 K(D, \partial _t D) \ d\Gamma dS _0 dt . \label{eq116h}
\end{equation}
The functional $I[D, \lambda_0, \lambda_1, \lambda_2, \lambda _n, \lambda _{P_\alpha}, \lambda _h]$ is formed from (\ref{eq116c}), (\ref{eq116d}), (\ref{eq116e}) and  (\ref{eq116h}):
\begin{eqnarray}
&& I[D, \lambda_0, \lambda_1, \lambda_2, \lambda _n, \lambda _{P_\alpha}, \lambda _h]  =  J[D, \lambda_0] - C_1[D, \lambda _1] - C_2[D, \lambda _2] \label{eq48} \cr\nonumber \\ 
&& - C_n[D, \lambda _n] - \sum_{\alpha = 1}^{3}C_{P_\alpha}[D, \lambda _{P_\alpha}] - C_h[D, \lambda _h] . \label{eq122}
\end{eqnarray} 
The existence of Lagrange multipliers $\lambda_0 \ne 0$, and $\lambda_1$, $\lambda_2 $, $\lambda _n$, $\lambda _{P_\alpha}$ ($\alpha = 1, 2, 3$) and $\lambda _h$ not all equal to zero, such that the variation of $I[D, \lambda_0, \lambda_1, \lambda_2, \lambda _n, \lambda _{P_\alpha}, \lambda _h]$ is stationary  $\delta I = 0$, represents a proof that it is possible to make $J[D]$ in (\ref{eq116f}) stationary subject to constraints (\ref{eq116}), (\ref{eq116a}) and (\ref{eq116b}).

For a function $D(x,p,t | (x_0,p_0)_\omega ,t_0 )$ to maximize $S_{I}^{DF}(t_a,t_0)$ subject to the constraints (\ref{eq116}), (\ref{eq116a}) and (\ref{eq116b}), it is necessary that it satisfies the Euler equation:
\begin{eqnarray}
&& \lambda _0 \left \{\frac{\partial K}{\partial D} - \frac{d}{dt}\left (\frac{\partial K}{\partial (\partial _t D)}\right ) - \sum_ {i = 1}^{3N} \left [\frac{d}{dx_i}\left (\frac{\partial K}{\partial (\partial _{x_i} D)}\right ) + \frac{d}{dp_i}\left (\frac{\partial K}{\partial (\partial _{p_i} D)}\right ) \right ]\right \} \cr\nonumber\\
&& - F\lambda _1  + F{\partial \lambda _2 \over \partial t} + F \int \left ({\partial \lambda _n \over \partial t}n + {\partial \lambda _h \over \partial t}h + \sum_{\alpha = 1}^3{\partial \lambda _{P_\alpha} \over \partial t} P_\alpha \right ) \, d^3{\bf r} \cr\nonumber\\
&& + F \int \left (\lambda _n \{n, H\} + \lambda _h \{h, H\} + \sum_{\alpha = 1}^3 \lambda _{P_\alpha} \{P_\alpha, H\}\right )\, d^3{\bf r}  = 0 \ .\label{eq123a}
\end{eqnarray}
The term multiplied by $\lambda _0$ in the Euler equation (\ref{eq123a}) is equal to zero, and from there it follows that
\begin{eqnarray}
&& {\partial \lambda _2 \over \partial t} + \int \left ({\partial \lambda _n \over \partial t}n + {\partial \lambda _h \over \partial t}h + \sum_{\alpha = 1}^3{\partial \lambda _{P_\alpha} \over \partial t} P_\alpha \right ) \, d^3{\bf r} \cr\nonumber\\
&& + \int \left (\lambda _n \{n, H\} + \lambda _h \{h, H\} + \sum_{\alpha = 1}^3 \lambda _{P_\alpha} \{P_\alpha, H\}\right )\, d^3{\bf r}  = \lambda _1 \ . \label{eq123b}
\end{eqnarray}
In this variational problem, the function $D(x,p,t | (x_0,p_0)_\omega ,t_0 )$ that renders $J[D]$ in (\ref{eq116f}) stationary subject to constraints (\ref{eq116}), (\ref{eq116a}) and (\ref{eq116b}), is not required to take on prescribed values on a portion of the boundary of integration region $M \times (t_0,t_a)$ in (\ref{eq116f}) where $t \neq t_0$. Therefore, in addition to satisfying the Euler equation (\ref{eq123a}), it is also necessary that it satisfies  the Euler boundary condition on the portion of the boundary of $M \times (t_0,t_a)$ where its values are not prescribed (ref. \cite{wan}). Accordingly, for all points on the portion of the boundary of $M \times (t_0,t_a)$ where $t = t_a$ the Euler boundary condition gives: 
\begin{eqnarray}
&& \left [\frac{\partial K}{\partial (\partial _t D)} - F\lambda _2 - F \int \left (\lambda _n n + \lambda _h h + \sum_{\alpha = 1}^3 \lambda _{P_\alpha} P_\alpha \right )\, d^3{\bf r}\right ] _{t = t_a}  \cr\nonumber\\ 
&& = - F\left [ \log D + \lambda _2 + \int \left (\lambda _n n + \lambda _h h + \sum_{\alpha = 1}^3 \lambda _{P_\alpha} P_\alpha \right )\, d^3{\bf r} \right ]_{t = t_a}  = 0 \ . \label{eq123c}
\end{eqnarray}
Furthermore, for all points on the portion of the boundary of $M \times (t_0,t_a)$ where time $t$ is in the interval $t_0 < t < t_a$, the Euler boundary condition gives:
\begin{equation}
F \left \{\left [\lambda _2 + \int \left (\lambda _n n + \lambda _h h + \sum_{\alpha = 1}^3 \lambda _{P_\alpha} P_\alpha \right )\, d^3{\bf r}\right ]{\bf v} \cdot {\bf n} \right \}_{\ \mathrm {at \ boundary \ of} \ M}  = 0 . \label{eq123d}
\end{equation}
In (\ref{eq123d}), ${\bf v} \cdot {\bf n}$ is a scalar product of the velocity vector field ${\bf v} (x,p)$ of points in $\Gamma $ (defined in \cite{kuic,kuic1}) and the unit normal ${\bf n}$ of the boundary surface of invariant set $M$, taken at the surface. Equation (\ref{eq123d}) is satisfied naturally due to Hamiltonian motion, since the set $M$ is invariant to Hamiltonian motion by definition, and therefore  ${\bf v} \cdot {\bf n} = 0$ for all points on the boundary surface of the invariant set $M$.

From the normalization condition (\ref{eq116}) and equation (\ref{eq123c}) we obtain the MaxEnt conditional probability density in the form:
\begin{equation}
D (x,p,t | (x_0,p_0)_\omega ,t_0 ) = \frac{1}{Z_t} \exp \left[ - \int d^3{\bf r} \left (\lambda _n n + \lambda _h h + \sum_{\alpha = 1}^3 \lambda _{P_\alpha} P_\alpha \right ) \right ] \ , \label{eq126} 
\end{equation}
where the Lagrange multiplier $\lambda _2((x_0,p_0)_\omega ,t_0; t) = \lambda _2 (t)$ is related to the normalization factor $Z(t)$ of the conditional probability density in the following way:
\begin{equation}
\lambda _2 (t) = \log Z(t) = \log \left \{\int_M d\Gamma \exp \left[ - \int d^3{\bf r} \left (\lambda _n n + \lambda _h h + \sum_{\alpha = 1}^3 \lambda _{P_\alpha} P_\alpha \right ) \right ]\right \} . \label{eq124} 
\end{equation}
In standard MaxEnt formalism the normalization factor $Z(t)$ is called the partition function, and in this case, we call it the partition functional because $Z(t) \equiv  Z_t \left [\lambda _n, \lambda _h, \lambda _{P_\alpha} \right ] \equiv  Z_t$. Furthermore, equation (\ref{eq124}) clearly shows why we need to keep the constraint (\ref{eq116a}) in this generalized approach; removing it would amount to putting the corresponding Lagrange multiplier $\lambda _2=0$, and thus introducing the excessive condition that $Z_t = 1$.

From the MaxEnt conditional probability density (\ref{eq126}) we obtain the microstate probability density:
\begin{equation}
f(x,p,t) = \frac{1}{Z_t} \exp \left[ - \int d^3{\bf r} \left (\lambda _n n + \lambda _h h + \sum_{\alpha = 1}^3 \lambda _{P_\alpha} P_\alpha \right ) \right ] \ . \label{eq127} 
\end{equation}
Equation (\ref{eq127}) is obtained with the help of (\ref{eq120}), using (\ref{eq118}), (\ref{eq126}) and the normalization condition of the path probability density $F((x_0,p_0)_\omega ,t_0)$. From (\ref{eq126}) and (\ref{eq127}) we notice immediately the following equality
\begin{equation}
D(x,p,t | (x_0,p_0)_\omega ,t_0 ) = f(x,p,t)\ . \label{eq128} 
\end{equation}
Also, from (\ref{eq126}) and (\ref{eq127}) it follows that at time $t$ the information entropy $S_I ^f (t)$ given by
\begin{equation}
S_{I}^{f} (t) = - \int_\Gamma f\log f \ d\Gamma , \label{eq128a}
\end{equation}
and the conditional information entropy $S_{I}^{DF}(t,t_0)$ given by  (\ref{eq115a}) are equal:
\begin{equation}
S_I ^f (t) = S_{I}^{DF}(t,t_0) = \log Z_t  + \int d^3{\bf r} \left (\lambda _n\langle n \rangle_t + \lambda _h \langle h \rangle_t + \sum_{\alpha = 1}^3 \lambda _{P_\alpha} \langle P_\alpha  \rangle_t \right ) \ .  \label{eq129} 
\end{equation}
From (\ref{eq128}), or (\ref{eq129}), it follows  that the initial phase space paths at time $t_0$ and final microstates at time $t$ are statistically independent, which as its logical consequence has a total loss of correlation. This result confirms the validity of the condition $t - t_0 \gg \tau $, where $\tau $ represents the time required for the loss of correlation between the initial phase space paths and final microstates. The reasons for the introduction of this condition are discussed in our previous papers \cite{kuic,kuic1} and in reference \cite{zubarev}.

The microstate probability density $f(x,p,t)$ given by (\ref{eq127}) is identical in form to the relevant distribution for the classical fluid in local equilibrium known from the literature \cite{zubarev}. With the assumption of local equilibrium, by simple comparison of the two distributions we obtain the following identifications of the Lagrange multipliers:
\begin{eqnarray}
\lambda _n({\bf r}, t) & = & - \beta ({\bf r}, t)\left ( \mu({\bf r}, t) - \frac{1}{2}m{\bf u}^2({\bf r}, t) \right )\nonumber\\
\lambda _{P_\alpha}({\bf r}, t)&  = & - \beta ({\bf r}, t)u_\alpha ({\bf r}, t)  \nonumber\\
\lambda _h({\bf r}, t)&  = & \beta ({\bf r}, t) \ . \label{eq130}  
\end{eqnarray}
In the reference \cite{zubarev} it is shown that $k^{-1}\beta ({\bf r}, t)^{-1} = T({\bf r}, t)$ has the role of local temperature, $\mu({\bf r}, t)$ of the local chemical potential per particle, and that ${\bf u}({\bf r}, t)$ is the velocity of the hydrodynamic motion. Furthermore, the identifications of the Lagrange multipliers $\lambda _n({\bf r}, t)$, $\lambda _{P_\alpha}({\bf r}, t)$ and $\lambda _h({\bf r}, t)$ given by (\ref{eq130}) show why we need all of the constraints given by the system of equations (\ref{eq116b}). Removing any of the constraints (\ref{eq116b}) would amount to putting the corresponding Lagrange multipliers equal to zero, and this would place excessive restrictions on the values of the local thermodynamic parameters $\beta ({\bf r}, t)$ and $\mu({\bf r}, t)$.

It is important also that the assumption of  local equilibrium gives a more precise physical definition of the condition  $t - t_0 \gg \tau $;  time $\tau $ required for the total loss of correlation between the initial phase space paths and final microstates is brought into relation with time $\tau _r$ required for the establishment of local equilibrium of the fluid with the relevant distribution given by (\ref{eq127}) and (\ref{eq130}).

Time derivative of the expression (\ref{eq124}) for $\log Z_t$, where $Z_t$ is the partition functional, gives 
\begin{equation}
{d \log Z_t \over dt} = \frac{1}{Z_t} {d Z_t \over dt} = - \int d^3{\bf r} \left ({\partial \lambda _n \over \partial t} \langle n \rangle_t + {\partial \lambda _h \over \partial t} \langle h \rangle_t + \sum_{\alpha = 1}^3 {\partial \lambda _{P_\alpha} \over \partial t} \langle P_\alpha \rangle_t \right ) \ . \label{eq131}  
\end{equation}
Time derivative of information entropy $S_I^{f}(t)$ in (\ref{eq129}) is obtained with the help of (\ref{eq131}), the constraints (\ref{eq116b}) and equations (\ref{eq105}) which are equivalent to these constraints,
\begin{eqnarray}
&& {d S_I ^f (t) \over dt} = \int d^3{\bf r} \left (\lambda _n {\partial \langle n \rangle_t \over \partial t} + \lambda _h {\partial \langle h \rangle_t \over \partial t} + \sum_{\alpha = 1}^3 \lambda _{P_\alpha} {\partial \langle P_\alpha \rangle_t \over \partial t}\right ) \cr\nonumber\\
&& = \int d^3{\bf r} \left (\lambda _n \langle \{n, H\} \rangle_t + \lambda _h \langle \{h, H\} \rangle_t  + \sum_{\alpha = 1}^3 \lambda _{P_\alpha} \langle \{P_\alpha, H\} \rangle_t \right ) \cr\nonumber\\
&& = - \int d^3{\bf r} \left (\lambda _n \nabla \cdot \langle \, {\bf J}\, \rangle_t + \lambda _h \nabla \cdot \langle {\bf J}_h\rangle_t + \sum_{\alpha = 1}^3   \lambda _{P_\alpha} \nabla  \cdot\langle {\bf J}_{P_\alpha}\rangle_t \right ) \ . \label{eq132}  
\end{eqnarray}
From the last line of (\ref{eq132}), we obtain that the time derivative of information entropy $S_I^{f}(t)$ is equal
\begin{eqnarray}
&& {d S_I ^f (t) \over dt}  = - \int d^3{\bf r} \left [\nabla \cdot (\lambda _n \langle \, {\bf J}\, \rangle_t ) + \nabla \cdot (\lambda _h \langle {\bf J}_h\rangle_t ) + \sum_{\alpha = 1}^3 \nabla \cdot (\lambda _{P_\alpha} \langle {\bf J}_{P_\alpha}\rangle_t ) \right ] \cr\nonumber\\
&& + \int d^3{\bf r} \left [\nabla (\lambda _n) \cdot \langle \, {\bf J}\, \rangle_t + \nabla (\lambda _h) \cdot \langle {\bf J}_h\rangle_t + \sum_{\alpha = 1}^3 \nabla
(\lambda _{P_\alpha}) \cdot\langle {\bf J}_{P_\alpha}\rangle_t \right ] \ . \label{eq133}  
\end{eqnarray}

By averaging the equation (\ref{eq123b}) over the microstate probability density $f(x,p,t)$ and using (\ref{eq124}), (\ref{eq131}) and (\ref{eq132}), it is easily shown that the Lagrange multiplier $\lambda _1$ depends only on time and that it is equal to the time derivative of information entropy,
\begin{eqnarray}
\lambda _1((x_0,p_0)_\omega ,t_0; t) = \lambda _1 (t) = {d S_I ^f (t) \over dt} \ . \label{eq135} 
\end{eqnarray}
Equation (\ref{eq135}) confirms the interpretation given to the Lagrange multiplier $\lambda _1$ in the basic model from the previous papers \cite{kuic,kuic1}. In the generalization of the approach, the introduction of  additional constraints (\ref{eq116b}), equivalent to the hydrodynamic continuity equations (\ref{eq105}), has determined precisely the rate of entropy change given by the  Lagrange multiplier $\lambda _1$; it is determined by (\ref{eq133}). It will now be shown that the obtained rate of entropy change (\ref{eq133}) is equal  to the corresponding standard expression from the thermodynamics of irreversible processes.

From the literature \cite{grandy,evans}, it is known that the current densities in the macroscopic conservation laws (\ref{eq105}) can be written in the following form:
\begin{eqnarray}  
m\langle {\bf J} ({\bf r}) \rangle_t &  = &\rho ({\bf r},t) {\bf u}({\bf r}, t) \ ,  \nonumber \\
\langle  J_{P_\alpha, \ \beta }({\bf r}) \rangle_t & = &\rho ({\bf r},t) u_\alpha ({\bf r}, t)u_\beta ({\bf r}, t) + T_{\beta \alpha }({\bf r}, t) \ ,  \nonumber \\
\langle J_{h, \ \alpha }  ({\bf r}) \rangle_t & = & \rho ({\bf r},t) e({\bf r},t)u_\alpha ({\bf r}, t) + T_{\alpha \beta}({\bf r}, t)u_\beta  ({\bf r}, t) + J_{Q, \  \alpha}({\bf r}, t) \ . \label{eq136} 
\end{eqnarray}
Here $\rho ({\bf r},t)$ is the mass density. The fluid velocity ${\bf u}({\bf r}, t)$ is the previously introduced velocity of hydrodynamic motion. In the second and third equation, ${\bf T}({\bf r}, t)$ is the pressure tensor  with the components $T_{\beta \alpha } ({\bf r}, t)$. With the help of Einstein summation convention for the indices $\alpha, \beta$ as introduced in the above equations, the pressure tensor is defined by the relation $dF_\alpha  \equiv - dS_\beta  T_{\beta \alpha }$, where $dF_\alpha  $ is the Cartesian component of the force $d{\bf F}$ across an infinitesimal surface element $d{\bf S}$. The pressure tensor is the negative of the stress tensor. In the third equation, $\rho ({\bf r},t) e({\bf r},t)$ is the energy density, where $e({\bf r},t)$ is the energy per unit mass. Heat current density is denoted by ${\bf J}_{Q}({\bf r}, t)$.

Densities of particle number, momentum and energy, which have previously been introduced in the macroscopic conservation laws (\ref{eq105}), are now analogously to (\ref{eq136}) written in the form
\begin{eqnarray}  
m\langle n ({\bf r})\rangle_t & = & \rho ({\bf r},t)\ ,  \nonumber \\
\langle {\bf P}({\bf r}) \rangle_t & = & \rho ({\bf r},t) {\bf u}({\bf r}, t)\ ,  \nonumber \\
\langle h ({\bf r})\rangle_t  & = & \rho ({\bf r},t) e({\bf r},t) \ . \label{eq137}
\end{eqnarray}
The fluid velocity ${\bf u}({\bf r}, t)$ can be consistently defined by the relation 
\begin{equation}
\langle {\bf J} ({\bf r}) \rangle_t = \langle n ({\bf r})\rangle_t {\bf u}({\bf r}, t) \ . \label{eq138}
\end{equation}
Using the identifications of Lagrange multipliers (\ref{eq130}) and relations (\ref{eq136}) for the current densities, from the last line of (\ref{eq132}) we obtain   
\begin{eqnarray}
&& {d S_I ^f (t) \over dt}  = - \int d^3{\bf r} \left \{- \beta \left ( \mu - \frac{1}{2}m{\bf u}^2 \right )\frac{1}{m}\nabla \cdot \rho  {\bf u} \right.  \cr\nonumber\\
&& \left. + \beta  \left [\nabla \cdot (\rho e {\bf u})  + \nabla \cdot \left ({\bf T}\cdot {\bf u}\right) + \nabla \cdot {\bf J}_{Q}\right ]  - \beta \left [ u_{\alpha }\nabla \cdot (\rho u_{\alpha} {\bf u}) + u_{\alpha } \partial_ {\beta} T_{\beta \alpha }\right ] \right \}  \ . \label{eq138a}
\end{eqnarray}
Here, Einstein summation convention is implied, and using it we can write 
\begin{eqnarray}
\nabla \cdot \left ({\bf T}\cdot {\bf u}\right) = u_{\alpha} \partial_{\beta} T_{\beta \alpha} + T_{\beta \alpha} \partial _{\beta} u_{\alpha} , \label{eq138b}
\end{eqnarray}
Furthermore, it follows also that
\begin{eqnarray}
u_{\alpha }\nabla \cdot (\rho u_{\alpha} {\bf u}) = \nabla \cdot \left (\rho \frac{1}{2}{\bf u}^2 {\bf u}\right ) + \frac{1}{2}{\bf u}^2\nabla \cdot (\rho {\bf u}) . \label{eq138c}
\end{eqnarray}
Then, using (\ref{eq138b}) and (\ref{eq138c}), from (\ref{eq138a}) we obtain
\begin{eqnarray}
&& {d S_I ^f (t) \over dt}  = - \int d^3{\bf r} \left \{- \beta  \mu \frac{1}{m}\nabla \cdot \rho  {\bf u} \right.  \cr\nonumber\\
&& \left. + \beta  \left [\nabla \cdot \left (\left (\rho e - \frac{1}{2}\rho {\bf u}^2 \right ){\bf u} \right )  +  T_{\beta \alpha} \partial _{\beta} u_{\alpha} + \nabla \cdot {\bf J}_{Q} \right ]  \right \}  \ . \label{eq138d}
\end{eqnarray}
Following reference \cite{evans}, the local thermodynamic internal energy density $U ({\bf r}, t)$ is obtained by removing the convective kinetic energy density from the total energy density:
\begin{eqnarray}
U ({\bf r}, t)= \rho e({\bf r}, t) - \frac{1}{2}\rho {\bf u}^2({\bf r}, t) . \label{eq138e}
\end{eqnarray}
Following \cite{evans}, we also define the viscous pressure tensor ${\bf \Pi}$ as the nonequilibrium part of the pressure tensor
\begin{eqnarray}
T_{\alpha \beta} ({\bf r}, t)= p({\bf r}, t) \delta _{\alpha \beta} + \Pi ({\bf r}, t) _{\alpha \beta} , \label{eq138f}
\end{eqnarray}
where $p({\bf r}, t)$ is the local pressure which comes from the assumption that the equilibrium equation of state is valid locally. Then, using (\ref{eq138e}) and (\ref{eq138f}) in  (\ref{eq138d}), it is easy to obtain
\begin{eqnarray}
&& {d S_I ^f (t) \over dt}  = - \int d^3{\bf r} \left \{ \nabla \cdot \left (\beta U {\bf u} \right )  +  \nabla \cdot \left (\beta  p {\bf u}\right) - \nabla \cdot \left (\beta  \mu \frac{1}{m}\rho  {\bf u}\right ) \right.  \cr\nonumber\\
&& \left. - U \nabla (\beta ) \cdot {\bf u}   -  \nabla \left (\beta  p \right) \cdot {\bf u} + \frac{1}{m}\rho  \nabla \left (\beta  \mu \right ) \cdot {\bf u} \right. \cr\nonumber\\
&& \left. + \beta  \left [\Pi _{\beta \alpha} \partial _{\beta} u_{\alpha} + \nabla \cdot {\bf J}_{Q} \right ]  \right \}  \ . \label{eq138h}
\end{eqnarray}
From the Euler equation for entropy,  
\begin{eqnarray}
S = \beta E + \beta pV -\beta \mu N , \label{eq138hh}
\end{eqnarray}
by applying it locally, using local extensive parameters $E$, $V$ and $N$, and then taking the quantities per unit volume, it follows that the local entropy density $s ({\bf r},t)$ is equal to
\begin{equation}
s ({\bf r},t)  =  \beta ({\bf r},t) U ({\bf r},t)  +  \beta  ({\bf r},t) p ({\bf r},t) - \beta ({\bf r},t) \mu ({\bf r},t) n ({\bf r},t) , \label{eq138i}
\end{equation}
where $n ({\bf r},t) = \langle n ({\bf r})\rangle _t = \rho ({\bf r},t) / m$ is the local particle-number density. From the Gibbs-Duhem relation
\begin{eqnarray}
E d\beta   +  Vd \left (\beta  p \right)  - N  d\left (\beta  \mu \right )  = 0 , \label{eq138j}
\end{eqnarray}
also applied locally and then written correspondingly for the local densities
\begin{eqnarray}
U d\beta   +  d \left (\beta  p \right)  - n  d\left (\beta  \mu \right )  = 0 , \label{eq138k}
\end{eqnarray}
it follows for the local time changes that
\begin{eqnarray}
U {\partial \beta \over \partial t}  +  {\partial \left (\beta  p \right)  \over \partial t} - n  {\partial \left (\beta  \mu \right ) \over \partial t}  = 0 . \label{eq138l}
\end{eqnarray}
Also, from (\ref{eq138k}) applied in the infinitesimal local system comoving with the fluid, it follows that
\begin{eqnarray}
U {\partial \beta \over \partial t}  +  {\partial \left (\beta  p \right)  \over \partial t} - n  {\partial \left (\beta  \mu \right ) \over \partial t}  + \left [U \nabla (\beta )  +  \nabla \left (\beta  p \right)  - n  \nabla \left (\beta  \mu \right )\right ] \cdot {\bf u} = 0 . \label{eq138m}
\end{eqnarray}
By comparing (\ref{eq138l}) and (\ref{eq138m}) we see that
\begin{eqnarray}
\left [U \nabla (\beta )  +  \nabla \left (\beta  p \right)  - n  \nabla \left (\beta  \mu \right )\right ] \cdot {\bf u} = 0 . \label{eq138n}
\end{eqnarray}
So, from (\ref{eq138h}), using (\ref{eq138i}) and (\ref{eq138n}), we obtain
\begin{eqnarray}
&& {d S_I ^f (t) \over dt}  = - \int_V d^3{\bf r} \left [ \nabla \cdot \left (s {\bf u} + \beta{\bf J}_{Q}\right )  \right ]  \cr\nonumber\\
&& + \int_V d^3{\bf r} \left [\nabla (\beta) \cdot {\bf J}_{Q} - \beta  \Pi _{\beta \alpha} \partial _{\beta} u_{\alpha} \right ] \ . \label{eq138o}
\end{eqnarray}
It is also important to note that, from the Euler equation (\ref{eq138i}), using the Gibbs-Duhem relation (\ref{eq138k}), we obtain for the local densities  
\begin{eqnarray}
ds = \beta dU -\beta \mu dn . \label{eq138p}
\end{eqnarray}
Relation (\ref{eq138p}) is in accordance with the approach developed by Callen \cite{callen}, where the entropy in a nonequilibrium system is defined locally, assuming the same dependence on the local extensive parameters as in equilibrium. Furthermore, the expression (\ref{eq138o}), when multiplied by the Boltzmann constant $k$, is consistent with the rate of entropy change for a single component classical fluid, that follows from the standard approach to nonequilibrium thermodynamics that assumes local equilibrium \cite{evans,deGroot}. By comparison with the references \cite{evans,deGroot}, we recognize that the divergence integral in (\ref{eq138o}) is the sum of convective and diffusive entropy flows over the boundary surface of the volume, and that the second integral in (\ref{eq138o}) is identical with the volume integral of the density of entropy production.

To simplify the calculation and identification of the aforementioned quantities, we introduce the new variables $(x^\prime,p^\prime)$ that are related with the old phase space variables $(x,p)$ by a canonical transformation which has the following form
\begin{eqnarray}
{\bf r}_k   =  {\bf r}_k^\prime \ , \qquad \quad {\bf p}_k  =  {\bf p}_k^\prime + m{\bf u}({\bf r}_k^\prime , t)  \ , \label{eq139}
\end{eqnarray}
where ${\bf u}({\bf r}_k^\prime , t)$ represents the fluid velocity at a position ${\bf r}_k^\prime$. It is easily checked that the Jacobian of this transformation is equal to unity. By applying the change of variables given by (\ref{eq139}), expressions (\ref{eq78}), (\ref{eq79}) and (\ref{eq80}) are transformed in the following way
\begin{eqnarray}
 n ({\bf r}) & = & n ^\prime ({\bf r})     \nonumber\\
 {\bf P}({\bf r}) & = & {\bf P}^\prime({\bf r}) + m{\bf u}({\bf r} , t) n ^\prime ({\bf r})  \nonumber\\
 h({\bf r}) & = & h^\prime({\bf r}) + {\bf u}({\bf r} , t) \cdot {\bf P}^\prime ({\bf r}) + \frac{1}{2}m{\bf u}^2({\bf r} , t) n ^\prime ({\bf r}) \ . \label{eq140}
\end{eqnarray}
Since the Jacobian of the transformation (\ref{eq139}) is equal to unity, the microstate probability density which in the new variables $(x^\prime,p^\prime)$ corresponds to the probability density (\ref{eq127}), is obtained by a simple introduction of (\ref{eq139}) and (\ref{eq140}):
\begin{eqnarray}
&& f^\prime (x^\prime, p^\prime, t) = \frac{1}{Z_t} \exp \left \{ - \int d^3{\bf r} \beta ({\bf r}, t)\left [h^\prime({\bf r}) - \mu ({\bf r}, t)n^\prime({\bf r}) \right ] \right \}  \ . \label{eq141} 
\end{eqnarray}
Because they are given by the integrals over the phase space, the information entropy $S_I^{f}(t)$ in (\ref{eq129}) and the corresponding partition function (\ref{eq124}) are invariant to the transformation (\ref{eq139}) for which the Jacobian of the transformation is equal to unity. With new coordinates $(x^\prime,p^\prime)$ explicitly indicated, the information entropy $S_I^{f}(t)$ is given by the expression
\begin{eqnarray}
S_I ^f (t) = \log Z_t + \int d^3{\bf r} \beta ({\bf r}, t)\left [\langle h^\prime({\bf r}) \rangle_t - \mu ({\bf r}, t)\langle n^\prime({\bf r}) \rangle_t \right ] \ ,  \label{eq142} 
\end{eqnarray}
while the partition function is given by the expression
\begin{eqnarray}
\log Z_t = \log \left \{\int_M d\Gamma ^{\prime} \exp \left[ - \int d^3{\bf r} \beta ({\bf r}, t)\left [h^\prime({\bf r}) - \mu ({\bf r}, t) n^\prime({\bf r}) \right ] \right ]\right \} \ . \label{eq143} 
\end{eqnarray}
By averaging the left side of (\ref{eq140}) over the microstate probability density given by (\ref{eq127}) and averaging the right side over the corresponding density (\ref{eq141}), and then using (\ref{eq137}), one obtains the following equalities
 \begin{eqnarray}
 \langle n ^\prime ({\bf r}) \rangle_t  & = & \langle n ({\bf r})\rangle_t    \nonumber\\
\langle {\bf P}^\prime({\bf r}) \rangle_t  & = & 0  \nonumber\\
 \langle h^\prime({\bf r}) \rangle_t  & = & \langle h({\bf r}) \rangle_t - \frac{1}{2}m{\bf u}^2({\bf r} , t) \langle n^\prime ({\bf r}) \rangle_t  \ . \label{eq144}
\end{eqnarray}
The last of equations (\ref{eq144}) is nothing but an expression for the internal energy density $U({\bf r} , t)$. Using the transformation (\ref{eq139}), and by assuming that the available part of phase space, the invariant set $M$ of all possible microstates, is invariant also on the transformation (\ref{eq139}), it is shown that the partition function (\ref{eq143}), probability density (\ref{eq141}) and the information entropy (\ref{eq142}) do not depend on the fluid velocity ${\bf u}({\bf r} , t)$. The time dependence in the expression for the information entropy $(\ref{eq142})$ appears only through the quantities $\beta ({\bf r}, t)$ and $\mu ({\bf r}, t)$, in an explicit and implicit way. For that reasons, the time derivatives of the information entropy (\ref{eq142}) also do not depend on ${\bf u}({\bf r} , t)$. Due to the invariance of the information entropy $S_I^{f}(t)$ on the transformation $(\ref{eq139})$, the same must be true also for the corresponding expressions (\ref{eq129}), (\ref{eq132}) and (\ref{eq133}). This means that, to simplify the calculations, everywhere in the expressions (\ref{eq129}), (\ref{eq130}), (\ref{eq132}) and  (\ref{eq133}) we can safely take that ${\bf u}({\bf r} , t) = 0$.

When this is done in (\ref{eq133}), using the expressions for the Lagrange multipliers (\ref{eq130}) and the current densities  (\ref{eq136}), or directly in (\ref{eq138o}), one obtains the corresponding expression for the time derivative of information entropy $S_I^{f}(t)$. By multiplying it with the Boltzmann constant $k$ and with the inclusion of $\beta ({\bf r}, t) = (kT({\bf r}, t))^{-1}$ we then obtain
\begin{eqnarray}
{d S (t) \over dt}  = - \int_V d^3{\bf r}\, \nabla \cdot \left (\frac{{\bf J}_Q({\bf r} , t)}{T({\bf r} , t)} \right )  + \int_V d^3{\bf r}\, \nabla \left (\frac{1}{T({\bf r} , t)}\right ) \cdot {\bf J}_Q({\bf r} , t)  \ . \label{eq146} 
\end{eqnarray}
Equation (\ref{eq146}) can also be recognized as the equation from the thermodynamics of irreversible processes that gives the {\it rate of change of entropy of the system}. The divergence integral in (\ref{eq146}) is the change due to entropy exchange through the boundary of the volume of the whole system, and accordingly, the sign in front of that term is negative. With regard to the initial assumption that the number of particles in the system is fixed, there is no exchange of particles with the environment; the given {\it system is closed}. Particles can not leave the volume of the system, so the fluid velocity vanishes at the boundary of the volume of the system; therefore, entropy  can not pass through this boundary by the streaming of the fluid. It is easy to see why the divergence integral in (\ref{eq146}) does not contain the contribution from the convective entropy flow present in (\ref{eq138o}); the convective entropy flow, if it is present within the closed system does not change the total entropy of the system, so its total contribution to the rate of entropy change is zero. Furthermore, we can also consider the limiting case in which the volume of the system is infinite. Then, also there is no exchange of heat with the environment, so the divergence integral in (\ref{eq146}) vanishes completely; that is the limit in which the system is isolated.   

The second integral in (\ref{eq146}) is the volume integral of the quantity known in the thermodynamics of irreversible processes as {\it the density of entropy production}, or alternatively, {\it entropy source strength}. One of the fundamental postulates of thermodynamics of irreversible processes \cite{evans,deGroot} is that this quantity always has the canonical form
\begin{equation}
\sigma = \sum_i {\bf X}_i \cdot {\bf J}_i\ . \label{eq147}
\end{equation}  
This canonical form defines the thermodynamic fluxes  ${\bf J}_i$  and the conjugate thermodynamic forces ${\bf X}_i$ denoted by the index $i$. The comparison of (\ref{eq147}) with the second integral in (\ref{eq146}) indicates that it is possible, by following this fundamental postulate, to define the density of entropy production for the classical fluid of identical particles considered here, in the form
\begin{equation}
\sigma ({\bf r} , t) = \nabla \left (\frac{1}{T({\bf r} , t)}\right ) \cdot {\bf J}_Q({\bf r} , t) \ . \label{eq148}
\end{equation}  
Proper identifications of the thermodynamic forces and fluxes are easily noticeable. From the comparison of the volume integrals of the density of entropy production given in (\ref{eq138o}) and in (\ref{eq146}), it follows that for a closed system,
\begin{equation}
\int_V d^3{\bf r}  (- T)^{-1}\partial _{\alpha } u_{\beta} \Pi _{\alpha \beta}  = 0 \ , \label{eq148a}
\end{equation}
where the Einstein summation convention for the components of the tensor $\partial _{\alpha } u_{\beta}$ and the viscous pressure tensor $\Pi _{\alpha \beta}$ is again implied. The entropy balance equation (\ref{eq146}) is valid also for a small volume comoving with the fluid with the local fluid velocity ${\bf u}({\bf r} , t)$. Therefore, the entropy production density is given only by (\ref{eq148}), and this means that locally
\begin{equation}
(- T)^{-1}\partial _{\alpha } u_{\beta} \Pi _{\alpha \beta}  = 0 \ . \label{eq148b}
\end{equation}
It is clear from the above arguments that (\ref{eq148b}) follows essentially from the requirement of local invariance of the entropy production density to the Galilean transformations, which is the standard requirement known from the literature \cite{deGroot}.

\section{Closed systems with external forcing} \label{cswef}

The results in the previous Section  were obtained in the setting applicable to the class of closed systems that are described by the Hamiltonian function that does not depend on time. This is applicable for systems that can exchange energy in the form of heat but can not exchange work and particles with the environment. The classical fluid of $N$ identical particles which is described by the Hamiltonian function (\ref{eq81}) is such a system. Further generalization to systems with Hamiltonian function that depends on time is straightforward. It was already explained in Sections III B, IV and V of the previous paper \cite{kuic1} and in Section \ref{secMIHIVE} of this paper. For example, if the Hamiltonian function (\ref{eq81}) also includes an additional term about which we have prior information, that describes the external time dependent potential field $\Phi _e({\bf r}, t)$,
\begin{equation}
H(x, p, t) = \sum_{i=1}^N  \left [\frac{{\bf p}_i^2}{2m} + \frac{1}{2}\sum_{j=1,\ j\neq i}^N  \Phi (\vert {\bf r}_i - {\bf r}_j \vert)\right ] + \sum_{i=1}^N \Phi _e({\bf r}_i , t) \ , \label{eq149c}
\end{equation}
then (\ref{eq149c}) describes the classical fluid of $N$ identical particles with the time dependent external force ${\bf F}_{e} ({\bf r}, t) = - \nabla \Phi _e({\bf r}, t)$ applied on it. This force may include also the effect of the walls of container confining the system of $N$ particles, if it can be described in such a way. Since the external potential $\Phi _e({\bf r}, t)$ has spatial and time dependence, for such a system of $N$ particles the total momentum and energy are not conserved and the local macroscopic conservation laws (\ref{eq105}) must be modified to include the effect of the external force ${\bf F}_{e} ({\bf r}, t)$:
\begin{eqnarray}  
\frac{\partial  \langle n ({\bf r})\rangle_t }{\partial t} & = & \langle \{n ({\bf r}), H\} \rangle_t = - \nabla \cdot \langle {\bf J} ({\bf r}) \rangle_t \ ,  \nonumber \\
\frac{\partial \langle P_\alpha  ({\bf r}) \rangle_t}{\partial t}   & = & \langle \{P_\alpha ({\bf r}), H\} \rangle_t =  - \nabla \cdot \langle {\bf J}_{P_\alpha}({\bf r}) \rangle_t + \langle n({\bf r})\rangle_t F_{e, \ \alpha } ({\bf r}, t)  \ ,  \nonumber \\
\frac{\partial \langle h ({\bf r})\rangle_t }{\partial t}    & = &  \langle \{h ({\bf r}), H\} \rangle_t =  - \nabla \cdot \langle {\bf J}_h ({\bf r}) \rangle_t + \langle n({\bf r})\rangle_t {\bf u}({\bf r}, t) \cdot {\bf F}_{e} ({\bf r}, t) \ . \label{eq149}
\end{eqnarray}
Here $\langle n({\bf r})\rangle_t{\bf F}_{e} ({\bf r}, t)$ is the external force times the local particle-number density,  i.e. the external force density. The right hand sides of (\ref{eq149}) including the external force terms are obtained from the Poisson brackets of the local dynamical variables $n ({\bf r})$, ${\bf P}({\bf r})$ and $h({\bf r})$ with the time dependent Hamiltonian function (\ref{eq149c}), averaged over the microstate probability density $f(x,p,t)$.  Therefore,  as shown in Sections \ref{secHJK} and \ref{secMIHIVE}, if the conditional probability density $D(x,p , t \vert (x_0,p_0)_\omega , t_0)$ is replaced by the conditional probability density $B(x,p , t \vert x_0,p_0 , t_0)$, path probability density $F((x_0,p_0)_\omega ,t_0 )$ with the microstate probability density $f(x_0,p_0, t_0)$, $M$ and $S_0(M)$ replaced by $ \Gamma $, and $dS_0$ replaced by $d\Gamma $, as is appropriate in the case of time dependent Hamiltonian function $H(x,p,t)$, then with all these replacements the constraints (\ref{eq116b}) are equivalent to (\ref{eq149}). Accordingly, as explained in Section \ref{secMIHIVE}, the same replacements are also done in the constraints (\ref{eq116}) and (\ref{eq116a}), and the conditional information entropy  $S_{I}^{DF}(t, t_0)$ is replaced by the conditional information entropy  $S_{I}^{Bf}(t, t_0)$. 

With all these replacements, and by applying the analogous maximization procedure to $S_{I}^{Bf}(t, t_0)$ as was applied to $S_{I}^{DF}(t, t_0)$ in Section \ref{secMIHIVE}, we obtain the MaxEnt conditional probability density $B(x,p , t \vert x_0,p_0 , t_0)$, which is analogous and of the same form as (\ref{eq126}) obtained for $D(x,p , t \vert (x_0,p_0)_\omega , t_0)$ in  Section \ref{secMIHIVE}. The only difference is with respect to the expressions for the time derivative of information entropy $S_I^{f}(t)$ in (\ref{eq132}) and (\ref{eq133}). Using (\ref{eq132}), (\ref{eq133}) and (\ref{eq149}) it is easy to see that the time derivative of information entropy $S_I^{f}(t)$ is given here by 
\begin{eqnarray}
&& {d S_I ^f (t) \over dt}  = - \int d^3{\bf r} \left [\nabla \cdot (\lambda _n \langle {\bf J}\rangle_t ) + \nabla \cdot (\lambda _h \langle {\bf J}_h \rangle_t ) + \sum_{\alpha = 1}^3 \nabla \cdot (\lambda _{P_\alpha}  \langle {\bf J}_{P_\alpha}\rangle_t ) \right ] \cr\nonumber\\
&& + \int d^3{\bf r} \left [\nabla (\lambda _n ) \cdot \langle \, {\bf J} \, \rangle_t + \nabla (\lambda _h ) \cdot \langle {\bf J}_h \rangle_t + \sum_{\alpha = 1}^3 \nabla
(\lambda _{P_\alpha}) \cdot\langle {\bf J}_{P_\alpha}\rangle_t \right ] \cr\nonumber\\
&& + \int d^3{\bf r} \left [\lambda _h \langle n \rangle_t {\bf u} \cdot {\bf F}_{e} + \sum_{\alpha = 1}^3   \lambda _{P_\alpha} \langle n \rangle_t F_{e, \ \alpha } \right ] \ . \label{eq150}  
\end{eqnarray}
The last line in (\ref{eq150}) is the contribution from the external force terms present in (\ref{eq149}). With the assumption of local equilibrium used here as in Section \ref{secMIHIVE}, all local thermodynamic identities used in Sec. \ref{secMIHIVE} are valid also here. The only difference is that along with the chemical part  $\mu_c ({\bf r}, t)$, the local chemical potential $\mu ({\bf r}, t)$ now also includes the external potential, i.e. $\mu ({\bf r}, t) = \mu_c ({\bf r}, t) + \Phi _e({\bf r}, t)$.  Using the identification of Lagrange multipliers (\ref{eq130}) and by following the same procedure as in Section \ref{secMIHIVE}, from (\ref{eq150}) here we also obtain the expression (\ref{eq138o}) for the time derivative of information entropy $S_I^f(t)$. Furthermore, in analogous way as in Section \ref{secMIHIVE}, we also obtain relations (\ref{eq146}), (\ref{eq148}),  (\ref{eq148a}) and (\ref{eq148b}). Further generalization to open systems is also straightforward and along with the derivation of the transport coefficients for the classical fluid it will be the subject of the further paper.

\section{Conclusion}
The construction of the probability distribution using the principle of maximum information entropy, i.e. by maximizing the information entropy subject to given constraints, includes in the probability distribution only the information which is represented by these constraints. The predictions derived from such probability distribution are the best predictions possible on the basis of available information, without the introduction of additional, uncertain assumptions. If control over certain macroscopic quantities reproduces some macroscopic phenomena in the experiment, then in accordance with the foundational principle of macroscopic reproducibility, the information about the values of these quantities is relevant for prediction of that macroscopic phenomena. Therefore, it can be said that consideration of the relevance of available information about the system for prediction and reproducibility of the macroscopic time evolution, is essential for a better understanding of the appearance of irreversibility. 

On the example of closed Hamiltonian system, it is shown that elementary description of irreversible macroscopic time evolution can be given, if the relevant information for nonequilibrium system on the chosen time scale is included in the probability distribution, by introducing it with the corresponding additional constraints on the maximization of the conditional information entropy. In this way, in the generalized approach developed in this paper, by introducing the hydrodynamic continuity equations as the relevant information on the hydrodynamic time scale, the rate of entropy change and the density of entropy production are obtained for the classical fluid of identical particles. The obtained expressions are in accordance with the definitions that these quantities have in the thermodynamics of irreversible processes. Therefore, as it is shown in this paper, even if our information about the microscopic dynamics is incomplete, the hydrodynamic continuity equations are, within the framework of predictive statistical mechanics developed in this paper, still sufficient for the description of irreversible macroscopic time evolution on the hydrodynamic time scale. It is also interesting in this context, as the group of authors have demonstrated \cite{gonzalez,davis}, that applying the principle of maximum information entropy under certain simple constraints can reproduce the laws of mechanics. If we consider that the predictive statistical mechanics is a general form of inference from available information without additional assumptions, based on the maximum information entropy principle and macroscopic reproducibility, the results obtained here suggest the importance of its basic principles for the theory of irreversibility.

\end{document}